\documentclass[screen]{acmart}

\usepackage{xcolor}
\usepackage{tabularx}

\AtBeginDocument{%
  \providecommand\BibTeX{{%
    \normalfont B\kern-0.5em{\scshape i\kern-0.25em b}\kern-0.8em\TeX}}}

\copyrightyear{2021}
\acmYear{2021}
\setcopyright{acmlicensed}\acmConference[RecSys '21]{Fifteenth ACM Conference on Recommender Systems}{September 27-October 1, 2021}{Amsterdam, Netherlands}
\acmBooktitle{Fifteenth ACM Conference on Recommender Systems (RecSys '21), September 27-October 1, 2021, Amsterdam, Netherlands}
\acmPrice{15.00}
\acmDOI{10.1145/3460231.3474241}
\acmISBN{978-1-4503-8458-2/21/09}

\usepackage{background}
\usepackage{stackengine}
\SetBgContents{This~article~has~been~accepted~for~publication~in~15th~ACM Conference on Recommender Systems. Association for Computing Machinery.~Citation~information:~https://doi.org/10.1145/3460231.3474241}

\SetBgScale{0.8}
\SetBgAngle{0}
\SetBgOpacity{1}
\SetBgColor{blue}
\SetBgHshift{0.1cm}
\SetBgVshift{15.5cm}

\begin{document}

\title
% short:
[An Audit of Misinformation Filter Bubbles on YouTube]
% full:
{An Audit of Misinformation Filter Bubbles on YouTube: Bubble Bursting and Recent Behavior Changes}
% OLD: Going Up the Rabbit hole: An Audit on Bursting Misinformation Bubbles on YouTube

%\author{Matus Tomlein, Branislav Pecher, Jakub Simko, Ivan Srba, Robert Moro, Elena Stefancova, Michal Kompan, Andrea Hrckova, Juraj Podrouzek, Maria Bielikova}
%\email{[name].[surname]@kinit.sk}
%\affiliation{%
%  \institution{Kempelen Institute of Intelligent Technologies}
%    \streetaddress{Mlynske nivy 5}
%  \city{Bratislava}
%  \country{Slovakia}
%}

\author{Matus Tomlein}
\affiliation{%
  \institution{\small Kempelen Institute of Intelligent Technologies \normalsize}
  \streetaddress{Mlynske nivy 5}
  \city{Bratislava}
  \country{Slovakia}
}
\email{matus.tomlein@kinit.sk}
\orcid{0000-0002-9960-700X}

\author{Branislav Pecher}
\affiliation{%
  \institution{\small Kempelen Institute of Intelligent Technologies \normalsize}
  \streetaddress{Mlynske nivy 5}
  \city{Bratislava}
  \country{Slovakia}
}
\email{branislav.pecher@kinit.sk}
\orcid{0000-0003-0344-8620}

\author{Jakub Simko}
\affiliation{%
  \institution{\small Kempelen Institute of Intelligent Technologies \normalsize}
  \streetaddress{Mlynske nivy 5}
  \city{Bratislava}
  \country{Slovakia}
}
\email{jakub.simko@kinit.sk}
\orcid{0000-0003-0239-4237}

\author{Ivan Srba}
\affiliation{%
  \institution{\small Kempelen Institute of Intelligent Technologies \normalsize}
  \streetaddress{Mlynske nivy 5}
  \city{Bratislava}
  \country{Slovakia}
}
\email{ivan.srba@kinit.sk}
\orcid{0000-0003-3511-5337}

\author{Robert Moro}
\affiliation{%
  \institution{\small Kempelen Institute of Intelligent Technologies \normalsize}
  \streetaddress{Mlynske nivy 5}
  \city{Bratislava}
  \country{Slovakia}
}
\email{robert.moro@kinit.sk}
\orcid{0000-0002-3052-8290}

\author{Elena Stefancova}
\affiliation{%
  \institution{\small Kempelen Institute of Intelligent Technologies \normalsize}
  \streetaddress{Mlynske nivy 5}
  \city{Bratislava}
  \country{Slovakia}
}
\email{elena.stefancova@kinit.sk}
\orcid{0000-0001-8683-939X}

\author{Michal Kompan}
\affiliation{%
  \institution{\small Kempelen Institute of Intelligent Technologies \normalsize}
  \streetaddress{Mlynske nivy 5}
  \city{Bratislava}
  \country{Slovakia}
}
\additionalaffiliation{%
  \institution{slovak.AI}
  \streetaddress{Ilkovicova 2}
  \city{Bratislava}
  \country{Slovakia}
}
\email{michal.kompan@kinit.sk}
\orcid{0000-0002-4649-5120}

\author{Andrea Hrckova}
\affiliation{%
  \institution{\small Kempelen Institute of Intelligent Technologies \normalsize}
  \streetaddress{Mlynske nivy 5}
  \city{Bratislava}
  \country{Slovakia}
}
\email{andrea.hrckova@kinit.sk}
\orcid{0000-0001-9312-6451}

\author{Juraj Podrouzek}
\affiliation{%
  \institution{\small Kempelen Institute of Intelligent Technologies \normalsize}
  \streetaddress{Mlynske nivy 5}
  \city{Bratislava}
  \country{Slovakia}
}
\email{juraj.podrouzek@kinit.sk}
\orcid{0000-0002-9691-0310}

\author{Maria Bielikova}
\affiliation{%
  \institution{\small Kempelen Institute of Intelligent Technologies \normalsize}
  \streetaddress{Mlynske nivy 5}
  \city{Bratislava}
  \country{Slovakia}
}
\additionalaffiliation{%
  \institution{slovak.AI}
  \streetaddress{Ilkovicova 2}
  \city{Bratislava}
  \country{Slovakia}
}
\email{maria.bielikova@kinit.sk}
\orcid{0000-0003-4105-3494}

\renewcommand{\shortauthors}{Tomlein, et al.}

\definecolor{DarkGreen}{RGB}{36,135,33}

\begin{abstract}
The negative effects of misinformation filter bubbles in adaptive systems have been known to researchers for some time. Several studies investigated, most prominently on YouTube, how fast a user can get into a misinformation filter bubble simply by selecting ``wrong choices'' from the items offered. Yet, no studies so far have investigated what it takes to ``burst the bubble'', i.e., revert the bubble enclosure. We present a study in which pre-programmed agents (acting as YouTube users) delve into misinformation filter bubbles by watching misinformation promoting content (for various topics). Then, by watching misinformation debunking content, the agents try to burst the bubbles and reach more balanced recommendation mixes. We recorded the search results and recommendations, which the agents encountered, and analyzed them for the presence of misinformation. Our key finding is that bursting of a filter bubble is possible, albeit it manifests differently from topic to topic. Moreover, we observe that filter bubbles do not truly appear in some situations. We also draw a direct comparison with a previous study. Sadly, we did not find much improvements in misinformation occurrences, despite recent pledges by YouTube.
\end{abstract}




% RECSYS: Authors will be asked to assign a selection of predefined custom tags to describe their paper in the submission system. Tags can be assigned to indicate algorithms, interfaces, automated or user-centric evaluations, for example. Reviewers will also report their expertise over these tags, and the information will be used in review assignments.

%  \includegraphics[width=\textwidth]{images/sampleteaser}

\maketitle

 \section{Introduction}

% SHORTER VERSION OF INTRODUCTION
% Suited more for RecSys

In this paper, we investigate the \emph{misinformation filter bubble} creation and bursting on YouTube. In our \emph{auditing study} we simulate user behavior on the YouTube platform, record platform responses (e.g., search results, recommendations) and manually annotate them for the presence of misinformative content. Then, we quantify the dynamics of misinformation filter bubble creation and also dynamics of bubble bursting, which is the novel aspect of the study. With this paper, we publish the implementation of the experimental infrastructure and also the data we collected\footnote{Available at \url{https://github.com/kinit-sk/yaudit-recsys-2021}}.

Our study adds to the previous works~\cite{Hussein2020, Papadamou2020, Spinelli2020, AbulFottouh2020, Ribeiro2020} that used \emph{audits} to quantify the portion of misinformative content being recommended on social media platforms. We directly build on works~\cite{Hussein2020, Papadamou2020, Spinelli2020} that observed and quantified the creation of misinformative filter bubbles on YouTube.

The general motivation of our work is to emphasize the \emph{need for independent oversight of personalization behavior of large platforms}. In the past, platforms have been accused of being contributors to the misinformation spreading due to their personalization routines. Simultaneously, they have been reluctant to revise these routines~\cite{zuboff2019age, vaidhyanathan2018antisocial}. And when they promise some changes, there is a lack of effective public oversight that could quantitatively evaluate their fulfillment. Auditing studies are tools that may improve such oversight.

While previous works investigated how a user can enter a filter bubble, no audits have covered \emph{if}, \emph{how} or with what \emph{``effort''} can the user ``burst'' (exit or lessen) the bubble. 
Multiple studies demonstrated that watching a series of misinformative videos would strengthen the further presence of such content in recommendations~\cite{Hussein2020, Papadamou2020, AbulFottouh2020}, or that following a path of the ``up next'' videos can bring the user to a very dubious content~\cite{Spinelli2020}. However, no studies investigated what type of user's watching behavior (e.g., switching to credible news videos or conspiracy debunking videos) would be needed to lessen the amount of misinformative content recommended to the user. 
Such knowledge would indeed be valuable. Not just for the sake of knowledge about the inner workings of YouTube's personalization, but also to improve the social, educational, or psychological strategies for building up resilience against misinformation.

\begin{quote}
    \it
    \textbf{As the first contribution}, this paper reports on the behavior of YouTube's personalization in a situation when a user with misinformation promoting watch history (i.e., with a developed misinformation filter bubble) starts to watch content debunking the misinformation (in an attempt to burst that misinformation filter bubble). The key finding is that watching misinformation debunking videos (e.g., credible news, scientific content) generally improves the situation (in terms of recommended items or search result personalization), albeit with varying effects and forms, mainly depending on particular misinformation topic.
\end{quote}

We aligned our methodology with previous works, most notably with the work of Hussein et al.~\cite{Hussein2020} who also investigated the creation of misinformation filter bubbles using user simulation. \emph{As part of our study, we replicated parts of Hussein's study}. We have done this for the sake of replication and to bootstrap bots with history of watching misinformation promoting videos. We re-used maximum of Hussein's seed data (topics, queries, videos), used similar scenarios and the same data annotation scheme. Therefore, we were able to directly compare the outcomes of both studies (e.g., on the number of observed misinformative videos present in recommendations or search results). Due to recent changes in YouTube policies~\cite{YouTube2020policies}, we expected to see less filter bubble creation behavior than Hussein et al. However, this was generally not the case.

\begin{quote}
    \it
    \textbf{As the second contribution}, we report changes in misinformation video occurrences on YouTube, which took place since the study of Hussein et al.~\cite{Hussein2020} (mid 2019). We observe worse situation regarding the topics of vaccination and (partially) 9/11 conspiracies and some improvements (less misinformation) for moon landing or chemtrails conspiracies.
\end{quote}

% Deliberately not putting in the paragraph on paper organization. 
\section{Background: Filter Bubbles and Misinformation}

To some extent, \emph{intellectual isolation} is a natural human defense against information overload~\cite{mutz2011communication} and provides us with stronger inner confidence~\cite{festinger1957theory}. However, it also comprises negative effects such as selective exposure (focusing on information that is in accordance with one’s worldview) or confirmation bias~\cite{del2016spreading, RePEc:now:jlqjps:100.00016037}. In social media, intellectual isolation contributes to the creation of \emph{echo chambers}~\cite{bessi2016personality}: the same ideas are repeated, mutually confirmed and amplified in relatively closed homogeneous groups. Polarization and fragmentation of the society increases~\cite{journals/corr/ZolloNVBMSCQ15,sunstein1999law}.

% filter bubbles

The negative effects of echo chambers can be amplified by \emph{filter bubbles}. Filter bubbles (as states of intellectual isolation) were firstly recognized by Pariser~\cite{pariser2011filter} as a negative consequence of personalization in social media and search engines. Researchers~\cite{pariser2011filter, sunstein1999law} agree that algorithms of such platforms support cognitive bias, as users are presented with the content that complies with their hitherto attitudes. Besides that, this effect also has ethical implications. Users are often unaware of the existence of filter bubbles, as well as of the information that was filtered out. Moreover, personalization and recommendation tailored to the users' interests can escalate the problems with misinformation~\cite{Spinelli2020}. 

% Misinformation part (written by Ivan)

\emph{Misinformation} is a false or inaccurate information that is spread regardless of an intention to deceive. Due to significant negative consequences of misinformation on our society (especially during the ongoing COVID-19 pandemic), tackling misinformation attracted a plethora of research efforts (see~\cite{Zannettou2019,Zhou2020} for recent surveys). While the majority of such research focuses on various characterization studies~\cite{simko2021study} or detection methods~\cite{pecher2021fireant,srba2019monant}, the studies investigating the relation between misinformation and adaptive systems are still relatively rare (e.g.,~\cite{Hussein2020,Papadamou2020}).

We denote filter bubbles that are characterized by the presence of misinformative content as \emph{misinformation filter bubbles}. They are states of intellectual isolation in false beliefs or a manipulated perceptions of reality. Analogically to \emph{topical} filter bubbles, misinformation filter bubbles can be characterized by a high homogeneity of recommendations/search results that share the same positive stance towards misinformation. In other words, the content adaptively presented to a user in a misinformation filter bubble supports one or several false claims/narratives. The proportion of such content represents how deep inside the bubble the user is.

To prevent misinformation and misinformation filter bubbles, social media conduct various countermeasures. These are usually reactions to public outcry or are required by legislation, e.g., EU's Code of practice on disinformation\footnote{\url{https://digital-strategy.ec.europa.eu/en/policies/code-practice-disinformation}}. Currently, the effectiveness of such countermeasures is evaluated mainly by self-evaluated reports. However, such reports are difficult to verify since social media are reluctant to provide access to their data for independent research.

The verification of countermeasures is further complicated by interference of psychological factors. For example, some researchers argue that cognitive bias is more influential than algorithms when it comes to intellectual isolation~\cite{del2016spreading, bakshy2015exposure}. To separate these influences, researchers employ platform \emph{audits}, such as the one in this paper.

% \todo[inline]{creating and bursting of a bubble}

% \todo[inline]{Initial state / state zero / vanilla recommendations -- which is different to the ``diversity'' state}

% \todo[inline]{Define the diversity (optimal, desired) state}
\section{Related work: Audits of Adaptive Systems}

In this context, an audit is a systematic statistical probing of an online platform, used to uncover socially problematic behavior underlying its algorithms~\cite{Sandvig2014Audits,Hussein2020}. Audits come in multiple forms~\cite{Sandvig2014Audits} and two of them are also suitable to investigate the effect of (misinformation) filter bubbles: \emph{crowdsourcing audits} and \emph{sockpuppeting audits}.

% crowdsourcing audits
Crowdsourcing audit studies are conducted using real user data. Silva et al.~\cite{Silva2020} developed a browser extension to collect personalized ads with real users on Facebook. Hannak et al.~\cite{Hannak2013} recruited Mechanical Turk users to run search queries and collected their personalized results. However, such auditing methodology suffers from a lack of isolation (users may be influenced by additional factors, e.g. confirmation bias). Moreover, uncontrolled environment makes comparisons difficult or unfeasible; it is difficult to keep users active; audits also raise several privacy issues.

% sockpuppeting audits
Sockpuppeting audits solve these problems by employing non-human bots that impersonate the behavior of users in a predefined controlled way~\cite{Sandvig2014Audits}. To achieve representative and meaningful results in sockpuppeting audits, researchers need to tackle several methodological challenges~\cite{Hussein2020}. First is the selection of appropriate seed data (e.g., the initial activity of bots, search queries). Second, the experimental setup must measure the real influence of the investigated phenomena. At the same time, it must minimize confounding factors and noise (e.g., of name, gender or geolocation~\cite{Hannak2013}). Another challenge is how to appropriately label the presence of the audited phenomena (expert-based/crowdsourced~\cite{Hussein2020,Silva2020} or automatic labeling~\cite{Papadamou2020} can be employed).  

% additional distinction features of audits
Audits can be further distinguished by the social media they are applied on (e.g., social networking sites~\cite{Silva2020,Papadamou2020,Hussein2020}, search engines~\cite{Metaxa2019,Le2019,Robertson2018}, e-commerce sites~\cite{Juneja2021}), by adaptive systems being investigated (e.g., recommendations~\cite{Hussein2020,Spinelli2020,Papadamou2020}, up-next recommendation~\cite{Hussein2020}, search results~\cite{Papadamou2020,Hussein2020,Le2019,Metaxa2019,Robertson2018}, autocomplete~\cite{Robertson2018}) and by phenomena being studied (e.g., misinformation~\cite{Hussein2020,Papadamou2020}, political bias~\cite{Le2019,Metaxa2019}, political ads~\cite{Silva2020}). In our study, we focus specifically on misinformation filter bubbles in the context of the online video platform YouTube and its recommender and search system. As argued by Spinelli et al.~\cite{Spinelli2020}, YouTube is an important case to study as a significant source of socially-generated content and because of its opaque recommendation policies.
% YouTube and YouTube RS
Some information about the inner workings of YouTube adaptive systems are provided by research papers published at RecSys conference~\cite{Covington2016,Zhao2019} or blogs~\cite{YouTube2020policies} published directly by the platform, nevertheless, a detailed information is unknown. Therefore, we feel a need to conduct independent auditing studies on undesired phenomena like unintended creation of misinformation filter bubbles.

% YouTube related audits
The existing studies confirmed the effects of filter bubbles in YouTube recommendations and search results. Spinelli et al.~\cite{Spinelli2020} found that chains of recommendations lead away from reliable sources and toward extreme and unscientific viewpoints. Similarly, Ribeiro et al.~\cite{Ribeiro2020} concluded that YouTube's recommendation contributes to further radicalization of users and found paths from large media channels to extreme content through recommendation. Abul-Fottouh et al.~\cite{AbulFottouh2020} confirmed a homophily effect in which anti-vaccine videos were more likely to recommend other anti-vaccine videos than pro-vaccine ones and vice versa.

% YouTube misinformation filter bubble audits
Recently, we can observe first audits focused specifically on misinformation filter bubbles. Hussein et al.~\cite{Hussein2020} and Papadomou et al.~\cite{Papadamou2020} found that YouTube mitigates pseudoscientific content in some handpicked topics such as COVID-19. Hussein et al.~\cite{Hussein2020} found that demographics and geolocation (within the US) affect personalization only after having acquired some watch history. These studies provide evidence of the existence and properties of misinformation filter bubbles on YouTube. From the properties that remain uninvestigated, we specifically address two. Firstly, the adaptive systems used by YouTube are in continuous development and improvement. Information on how YouTube proceeds in countering misinformation is needed. Secondly, while the existing studies focused on misinformation filter bubble creation, we do not have the same perspective on the inverse process -- filter bubble bursting.
\section{Study design and methodology}

To investigate the dynamics of bursting out of a misinformation filter bubble, we conducted an agent-based  sockpuppeting audit study. The study took place on YouTube, but its methodology and implementation can be generalized to any adaptive service, where recommendations can be user-observed.

In the study, we let a series of agents (bots) pose as YouTube users. The agents performed pre-defined sequences of video watches and query searches. They also recorded items they saw: recommended videos and search results. The pre-defined actions were designed to first \emph{invoke the misinformation filter bubble effect} by purposefully watching videos with (or leaning towards) misinformative content. Then, agents tried to \emph{mitigate the bubble effect} by watching videos with trustworthy (misinformation debunking) content. Between their actions, the agents were idle for some time to prevent possible carry-over effects. The degree of how deep inside a bubble the agent is was observed through the number and rank of misinformative videos offered to them.

The secondary outcome is the partial replication of a previous study done by Hussein et al.~\cite{Hussein2020} (denoted onwards as the \emph{reference study}). This replication allowed us to draw direct comparisons between quantities of misinformative content that agents encountered now (March 2021) and during the reference study done in mid 2019.

\subsection{Research Questions, Hypotheses and Metrics}
\label{ssec:research_questions}

\textbf{RQ1 (comparison to the reference study):} \emph{Has YouTube's personalization behavior changed with regards to misinformative videos since the reference study?} In particular, we seek to validate the following hypothesis:
\begin{itemize}
    \item \textbf{H1.1:} Compared on \emph{SERP-MS} and \emph{normalized score} metrics (see below), we would see better scores (after constructing a promoting watch history) than in the reference study in both search and recommendations (given YouTube's pledges~\cite{YouTube2020policies}).
\end{itemize}

\textbf{RQ2 (bubble bursting dynamics):} \emph{How does the effect of misinformation filter bubbles change, when debunking videos are watched?} The ``means of bubble bursting'' would be implicit user feedback -- watching misinformation debunking videos. In particular, we seek to validate the following hypotheses:
\begin{itemize}
    \item \textbf{H2.0:} Watching videos belonging to promoting misinformation stance leads to their increased presence in both search results and recommendations (worse SERP-MS and normalized score metrics).
    \item \textbf{H2.1:} Watching the sequence of misinformation debunking videos after the sequence of misinformation promoting videos will improve the metrics \emph{in comparison to the end of the promoting sequence}.
    \item \textbf{H2.2:} Watching the sequence of misinformation debunking videos after the sequence of misinformation promoting videos will improve the metrics \emph{in comparison to the start of the experiment}.
\end{itemize}

The metrics we use -- \emph{SERP-MS} and \emph{normalized score} -- are drawn directly from the reference study. Both metrics quantify misinformation prevalence in a given list of items (videos), which are annotated as either \emph{promoting} (value 1), \emph{debunking} (value -1) or \emph{neutral} (value 0). The output of both metrics is, similarly, from the $\left \langle -1, 1  \right \rangle$ interval. Lists populated mostly with debunking content would receive values close to -1, with promoting close to 1 and with balanced or mostly neutral, close to 0. In other words, a score closer to -1 means better score.
\begin{description}
    \item [Normalized score.] A metric computed as average of individual annotations of items present in the list. It is suited for unordered, shorter lists (in our case, recommendations).
    \item [SERP-MS (Search result page misinformation score).] A metric capturing amount of misinformation and its rank. It is suited for longer, ordered lists (in our case, search results). It is computed as $SERP\mbox{-}MS = \frac{\sum_{r=1}^{n}(x_i*(n-r+1))}{\frac{n*(n+1)}{2}}$, where $x_i$ is annotation value, $r$ search result rank and $n$ number of search results in the list~\cite{Hussein2020}.
\end{description}

\subsection{Experiments scenarios}

\begin{figure*}
\centering
\includegraphics[width=\textwidth]{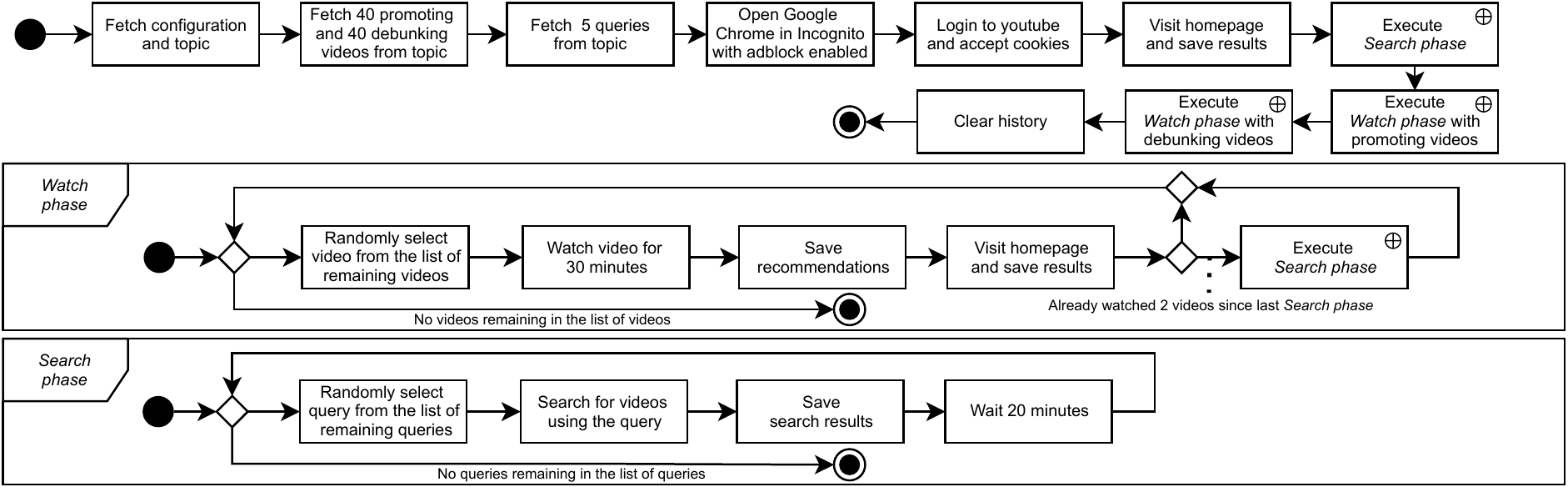}
\caption{Agent scenario for creating and bursting misinformation filter bubbles}
\label{fig:agent-scenario}
\Description[The figure illustrates the bot scenario described in text]{The figure illustrates the four phases of the bot scenario, as is described in the text, in a form of diagram.}
\end{figure*}

We let agents interact with YouTube following a \emph{scenario} composed of four phases, as depicted in Figure \ref{fig:agent-scenario}.

% JS: algorithm moved to unused text file, was too long :(

\emph{Phase 0: Agent initialization.} At the start of a run, the agent fetches its desired configuration, including the YouTube user account and various controlled variables (the variable values are explained further below). Also, the agent fetches $\tau \in T$, a topic with which it will work (e.g., ``9/11''). The agent fetches $V_{prom}$ and $V_{deb}$, which are lists of $n_{prom}=40$ and $n_{deb}=40$ most popular videos promoting, respectively debunking, misinformation within topic $\tau$. Afterward, it fetches $Q$, a set of $n_{q}=5$ search queries related to the particular $\tau$ (e.g., ``9/11 conspiracy``). The agent configures and opens a browser in incognito mode, visits YouTube, logs in using the given user account, and accepts cookies. Finally, the agent creates a neutral baseline by visiting the homepage and saving videos, and performing a search phase. In the \emph{search phase}, the agent randomly iterates through search queries in $Q$, executes each query on YouTube, and saves the search results. To prevent any carry-over effect between search queries, the agent waits for $t_{wait}=20$ minutes after each query.

\emph{Phase 1 (promoting): Create the filter bubble.} For creating a filter bubble effect, the agent randomly iterates through $V_{prom}$ and ``watches'' each video for $t_{watch}=30$ minutes (or less, if the video is shorter). Immediately after watching a video, the agent saves video recommendations on that video's page and visits the YouTube homepage, saving video recommendations listed there as well. After every $f_{q}=2$ videos, the agent performs another search phase.

\emph{Phase 2 (debunking): Burst the filter bubble.} The agent follows the same steps as in phase 2. The only difference is the use of $V_{deb}$ instead of $V_{prom}$. 

\emph{Phase 3: Tear-down.} In this phase, the agent clears YouTube history (using Google's ``my activity`` section), making the used user account ready for the next run.

For each selected topic, we run the scenario 10 times (in parallel). This way, we were able to deal with recommendation noise present at the platform. In order to run our experiments multiple times, we used the \emph{reset} (delete all history) button provided by Google instead of creating a new user profile for each run. %We delete the whole history in one go, not only from the YouTube platform but also from all other Google products, such as Google Ads and Google Search.
Before deciding to use the \emph{reset} button in our study, we first performed a short verification study to see whether using this button really deletes the whole history and resets the personalization on YouTube. We randomly selected few topics, from which we manually watched few videos (5 for each). Then, we used the reset button and evaluated the difference between videos appearing on the YouTube homepage, recommendations, and search. We found no carry-over effects.

%\subsubsection{Controlled variables and parameter setup pre-study}

% The comments here are just for reference

% $n_q$ - number of search queries
% $Q$ - set of queries
% $\tau$ - topic selected for the run
% $T$ - list of topics selected for experiments

% $V$ - list of all videos
% $V_{prom}$ - list of promoting videos
% $V_{deb}$ - list of debunking videos

% $n_{prom}$ - number of promoting videos
% $n_{deb}$ - number of debunking videos

% $t_{wait}$ - waiting time after querying
% $t_{watch}$ - watching time for one video

% $f_{q}$ - number of videos to watch before using search queries

We needed to set up several attributes of agents (e.g., YouTube user profiles). For \emph{geolocation}, we use N.~Virginia to allow for better comparison with the reference study. The date of birth for all accounts was arbitrarily set to 6.6.1990 to represent a person roughly 30 years old. The gender was set as ``rather not say'' to prevent any personalization based on gender. The names chosen for the accounts were composed randomly of the most common surnames and unisex given names used in the US.

There were also \emph{process parameters} that we needed to keep constant. These include 1) $n_{prom}=40$ and $n_{deb}=40$ representing the number of seed videos used in promoting and debunking phases; 2) $t_{watch}=30$ representing the maximum watching time in minutes for every video; 3) $n_{q}=5$ representing the number of queries used; 4) $t_{wait}=20$ representing the wait time in minutes between query yields and 5) $f_{q}=2$ representing the number of videos to watch between search phases. 

Values of the \emph{process parameters} greatly influence the total running time and results of the experiment. Yet, determining them was not straightforward given many unknown properties of the environment (first and foremost YouTube's algorithms). For example, prior to the experiment, it was unclear how often we need to probe for changes in recommendations and search result personalizations to answer our research questions.

Therefore, \emph{we run a pre-study in which we determined the best parameter setup}. Measuring the Levenshtein distance between ordered results and overlap of lists of recommended videos we determined to run 10 individual agents for each topic, as we observed instability between repeated runs (e.g., the same configuration yielded \(\sim70\%\) of the same recommended videos). For the $n_{prom}$ and $n_{deb}$ parameters, we observed that in some cases, a filter bubble could be detected after 20 watched videos. Yet in others, it was 30 or more. Due to this inconsistency, we opted to watch 40 videos for a phase. To determine the optimal value of $t_{watch}$, we first calculated the average running time of our seed videos. Most of the videos (\(\sim85\%\)) had a running time of about 30 minutes or shorter, so 30 minutes became the baseline value. In addition, we compared the results obtained by watching only 30 minutes with results from watching the whole video regardless of its length, but found no apparent differences. 

To determine the number of queries $n_{q}$ and periodicity of searches $f_{q}$, we ran the scenario with all seed queries introduced by the reference study and used them after every seed video. We observed that the difference in search results between successive seed videos was not significant. As the choice of search queries and the frequency of their use greatly prolonged the overall running time of the agents, we opted to run the search phase after every second video. In addition, we opted to use only 5 queries per topic.

The only parameter not set by a pre-study is $t_{wait}$, which we set to 20 minutes based on previous studies. These found that the carry-over effect (which we wanted to avoid) is visible for 11 minutes after the search~\cite{Hannak2013, Hussein2020}.

\subsection{Seed Data}

We used 5 topics in our study (same as the reference study): 1) \emph{9/11 conspiracies} claiming that authorities either knew about (or orchestrated) the attack, or that the fall of the twin towers was a result of a controlled demolition, 2)\emph{moon landing conspiracies} claiming the landing was staged by NASA and in reality did not happen, 3) \emph{chemtrails conspiracy} claiming that the trails behind aircraft are purposefully composed of dangerous chemicals, 4) \emph{flat earth conspiracy} claiming that we are being lied to about the spherical nature of Earth and 5) \emph{vaccines conspiracy} claiming that vaccines are harmful, causing various range of diseases, such as autism. The narratives associated with the topics are \emph{popular} (persistently discussed), while at the same time, \emph{demonstrably false}, as determined by the reference study~\cite{Hussein2020}.

For each topic, the experiment required two sets of seed videos. The \emph{promoting} set, used to construct a misinformation filter bubble (its videos have a promoting stance towards the conspiratorial narrative or present misinformation). And the \emph{debunking} set, aimed to burst the bubble (and contains videos disproving the conspiratorial narratives).

As a basis for our seed data sets we used data already published in the reference study, which the authors either used as seed data, or collected and annotated. To make sure we use adequate seed data, we re-annotated all of them.

The number of seed videos collected this way was insufficient for some topics (we required twice as many seed videos as the reference study). To collect more, we used an extended version of the seed video identification methodology of the reference study. Following is the list of approaches we used (in a descending order of priority): YouTube search, other search engines (Google search, Bing video search, Yahoo video search), YouTube channel references, recommendations, YouTube homepage, and known misinformation websites. To minimize any biases, we used a maximum of 3 videos from the same channel. 

As for search queries, we required fewer of them than the reference study. We selected a subset based on their popularity on YouTube. Some examples of the used queries are: "\emph{9/11 conspiracy}", "\emph{Chemtrails}", "\emph{flat earth proof}", "\emph{anti vaccination}", "\emph{moon landing fake}".

\subsection{Data collection and annotation}
\label{sec:annotation-methodology}

Agents collect videos from three main components on YouTube: 1) \emph{recommendations} appearing next to videos presently watched, 2) \emph{home page} videos and 3) \emph{search results}. In case of recommendations, we collect 20 videos that YouTube normally displays next to a currently watched video (in rare cases, less than 20 videos are recommended). For home page videos and search results, we collect all videos appearing with the given resolution, but no less than 20. In case when less than 20 videos appear, the agent scrolled further down on the page to load more videos.

For each video encountered, the agent collects metadata: 1) \emph{YouTube video ID}, 2) \emph{position} of the video in the list, and 3) \emph{presence of a warning/clarification message} that appears with problematic topics such as COVID-19. Other metadata, such as video \emph{title}, \emph{channel} or \emph{description}, are collected using the YouTube API.

To annotate the collected videos for the presence of misinformation, we used an extended version of the methodology proposed in the reference study. Each video was viewed and annotated by the authors of this study using a code ranging from -1 to 10. The videos are annotated as \emph{debunking} (code -1), when their narrative provides arguments against the misinformation related to the particular topic (such as \emph{"The Side Effects of Vaccines - How High is the Risk?"}), \emph{neutral} (code 0) when the narrative discusses the related misinformation but does not present a stance towards it (such as \emph{"Flat Earthers vs Scientists: Can We Trust Science? | Middle Ground"}), and \emph{promoting} (code 1), when the narrative promotes the related misinformation (such as \emph{"MIND BLOWING CONSPIRACY THEORIES"}). The codes 2, 3, and 4 have the same meaning as codes -1, 0, and 1, but are used in cases when they discuss misinformation not related to the topic of the run (e.g., video dealing with climate crisis misinformation encountered during a flat earth audit). The code 5 is applied to videos that do not contain any misinformation views (such as \emph{"Gordon's Guide To Bacon"}). This includes completely unrelated videos (e.g., music or reality show videos), but also videos that are related to the general audit topic, but not misinformation (e.g., original news coverage of 9/11 events). In rare cases of videos that are not in English and do not provide English subtitles, code 6 is assigned. Also rare are the cases when the narrative of the video cannot be determined with enough confidence (code 7). Videos removed from YouTube (before they are annotated) are coded as 8. Finally, as an extension of the approach used in the reference study, we use codes 9 and 10 to denote videos that specifically mention misinformation but rather than debunk them, they mock them (9 for related misinformation, 10 for unrelated misinformation, for example \emph{"The Most Deluded Flat Earther in Existence!"}). Mocking videos are a distinct (and often popular) category, which we wanted to investigate separately (however, for the purposes of analysis, they are treated as debunking videos).

To determine how many annotators are needed per video, we first re-annotated the seed videos released by the reference study. Each was annotated by at least two authors, and the annotations were compared between each other and with annotations from the reference study. We achieved Cohen's kappa value of 0.815 between us and 0.688 with the reference study. We identified characteristics of edge cases. Following the re-annotation and the findings from it, when annotating our collected videos, we assign only one annotator per collected video with instructions to indicate and comment if an edge case video is encountered. These were then reviewed by another annotator.

For the purpose of this study and to evaluate our hypotheses, we annotated the following subset of collected videos:
\begin{itemize}
    \item All recorded \emph{search results}.
    \item Videos recommended for first 2 seed videos at the start of the run and last 2 seed videos of both phases (resulting in 6 sets of annotated videos per topic). This selection was a compromise between representativeness, correspondence to the reference study, and our capacities.
    \item We have \emph{not} annotated the \emph{home page videos} for the purpose of this study. These videos were the most numerous, the most heterogeneous, and with little overlap across bots and seed videos.
\end{itemize}

\subsection{Data ethics assessment}

To consider various ethical issues regarding the research of misinformative content, we carried out a series of data ethics workshops. We explored questions related to data ethics issues~\cite{Tranberg2020Dataethics} within our audit and its impact on stakeholders. Based on the topics that emerged during the data ethics workshops, we identified different stakeholder groups. The most affected ones were platform users, annotators, content creators, and other researchers. For every stakeholder group, we devised different engagement strategies and specific action steps. Our main task was to devise countermeasures to the most prominent risks that could emerge for these stakeholder groups.

First, we were concerned about the risk of unjustified flagging of the content as misinformation and their creators as conspirators. To minimize this risk, we decided to report hesitations in the annotation process. These hesitations were consequently back-checked by other annotators and independently validated until the consensus was reached. One of our main concerns was also not to harm or delude other users of the platform. To avoid disproportional boost of the misinformation content by our activity, we select the videos with at least 1000 views and warn annotators not to watch videos online more than one time, or in case of back-checks, two times. After each round, we reset user account and delete the watch history. 

%We were also aware of several possible biases considering the dynamics of bubble creating and bubble bursting for various groups of users and different environments. Our results are tied to the specific type of users, platform and misinformative content.  To validate our results on a more diverse set of users is a future challenge. 

Other concerns were connected to the deterioration of well-being of human annotators. Specifically, that their decision-making abilities would be negatively affected after a long annotation process. We proposed the daily routines for annotation, including the breaks during the process and advised to monitor any changes in annotators beliefs. Our annotators also underwent the survey on their tendency to believe in conspiracy theories\footnote{\url{https://openpsychometrics.org/tests/GCBS/}} and none of them showed such tendency at the end of the study.

%To achieve reproducibility of our research and to support any other researchers in this area, we publish the source code and data sets. 

% among other things: we recognize that bots may be harmful
% ? we also consider terms and conditions unfair
% check terms and conditions

% Moved from section 5
\subsection{A note on comparability with the reference study by Hussein et al.}
\label{comparison-explained}

In order to be able to draw comparisons, we kept the methodology of our study as compatible as possible with the previous study by Hussein et al.~\cite{Hussein2020}. We shared the general approach of prompting YouTube with implicit feedback: both studies used similar scenarios of watching a series of misinformation promoting videos and recording search results and recommended videos. We re-used the topics, a subset (for scaling reasons) of search queries, and all available seed videos (complementing the rest by using a similar approach as the reference study). Moreover, both studies used the same coding scheme, metrics, sleep times, and annotated a similar number of videos. 

We should also note differences between the studies, which mainly source from different original motivations for our study. For instance, no significant effects of demographics and geolocation of the agents were found in the reference study, so we only controlled these. In Hussein's experiments, all videos were first ``watched'' and only then all search queries were fired. In our study, we fired all queries after watching \emph{every 2nd} video (with the motivation to get data from the entire run, not just the start and end moment). The reference study created genuine 150 accounts on YouTube, while we used fewer accounts and took advantage of the browsing history reset option. In some aspects, our study had a larger scale: we executed 10 runs for each topic instead of one (to reduce possible noise) and used twice as many seed videos (to make sure that filter bubbles develop). There were also technical differences between the setups, as we used our own implementation of agents (e.g., different browser, ad-blocking software).

Given the methodological alignment (and despite the differences), we are confident to directly compare some of the outcomes of both studies, namely quantity of misinformative content appearing at the end of the promoting phases.
\section{Results and findings}

Following the study design, we executed the study between March 2nd and March 31st, 2021. Together, we executed 50 bot runs (10 for each topic). On average, runs for a single topic took 5 days (bots for a topic ran in parallel). The bots watched 3951 videos (collected 78763 recommendations associated with them, 8526 of them unique), executed 10075 queries (collected 201404 search results, 942 of them unique), and visited homepage 3990 times (collected 116479 videos there, 9977 of them unique). Overall, we recorded 17405 unique videos originating from 6342 channels.

Using the selection strategy and annotation scheme described in Section~\ref{sec:annotation-methodology}, 5 annotators annotated unique 2914 videos (covering 255844 appearances). In total, 244 videos were identified as promoting misinformation (related or unrelated to respective topics), 628 as debunking (including mocking videos), 184 as neutral, 1829 as not about misinformation. Other videos (unknown, non-English, or removed) numbered 29.

We report the results according to research questions and hypotheses defined in Section~\ref{ssec:research_questions}.
SERP-MS score metrics are reported for search results and mean normalized scores for recommendations.
Since the metrics are not normally distributed with some samples of unequal sizes, we make use of non-parametric statistical tests.
Pairwise tests are performed using two-sided Mann-Whitney U test.
In cases where multiple comparisons by topics are performed, Bonferroni correction is applied on the significance level (in that case \(\alpha=0.05\) is divided by number of topics \(n_T=5\), resulting in \(\alpha=0.01\)).

% note on compatibility with reference study moved to the end of section 4

\subsection{RQ1: Has YouTube's personalization behavior changed since the reference study?}

%%% Overall results across topics
Overall, we see a small change in the mean SERP-MS score across the same search queries in our and reference data: mean SERP-MS worsened from -0.46 (std 0.42) in reference data to -0.42 mean (std 0.3) in our data.
However, the distributions are not statistically significantly different (n.s.d.).
There is a similar small change towards the promoting spectrum in up-next (first result in recommendation list) and top-5 recommendations (following 5 recommendations).
We compared the up-next and top-5 recommendations together (as top-6 recommendations) using last 10 watched promoting videos in reference watch experiments and last two watched videos in our promoting phase.
We see mean normalized score worsened from -0.07 (std 0.27) in reference data to -0.04 (std 0.31) in our data.
These distributions are also not significantly different (U=45781.5, n.s.d.).% and p-value=0.3045).

%%% Results within topics
More considerable shifts in the data can be observed when looking at individual topics.
Table~\ref{tab:hussein-comparison-search} shows a comparison of SERP-MS scores for top-10 search results between our and reference data.
Improvement can be seen within certain queries for the chemtrails conspiracy that show a large decrease in the number of promoting videos.
The reference study reported that this topic receives significantly more misinformative search results compared to all other topics.
In our experiments, their proportion was lower than in the 9/11 conspiracy.
On the other hand, search results for flat earth conspiracy worsened.
Queries such as ``flat earth british'' resulted in more promoting videos, likely due to new content on channels with similar names.
Within the anti-vaccination topic, there is an increase in neutral videos (from 12\% to 35\%) and thus a drop in debunking videos (from 85\% to 61\%).
This may relate to new content regarding COVID-19.

Table~\ref{tab:hussein-comparison-recommendations} shows a comparison of normalized scores for up-next and top-5 recommendations.
Only the moon landing and anti-vaccination topics come from statistically significantly different distributions.
Similar to search results, recommendations for the 9/11 and anti-vaccination conspiracy topics worsened.
There were more promoting videos on the 9/11 topic (27\% instead of 18\%).
In the anti-vaccination topic, we observed a drop in debunking videos (from 29\% to 9\%) and a subsequent increase in neutral (from 70\% to 78\%) and promoting videos (from 1\% to 8\%).
The change within the anti-vaccination controversy is even more pronounced when looking at up-next recommendations separately.
Within up-next, the proportion of debunking videos drops from 77\% to 19\%, neutral videos increase from 22\% to 70\%, and promoting increase from 1 to 11\%.
On the other hand, in the moon landing topic, we see much more debunking video recommendations---40\% instead of 23\% in reference data.

%%% Interpretation of results
These results bring up a need to distinguish between \emph{endogenous} (changes in algorithms, policy decisions made by platforms to hide certain content) and \emph{exogenous} factors (changes in content, external events, behavior of content creators) as discussed by Metaxa et al.~\cite{Metaxa2019}.
Our observations show that search results and recommendations were in part influenced by exogenous changes in content on YouTube.
Within the chemtrails conspiracy, we observed results related to a new song by Lana del Rey that mentions ``Chemtrails'' in its name.
Search results and recommendations in the anti-vaccination topic seem to be influenced by COVID-19.
Flat earth conspiracy videos were influenced by an increased amount of activity within a single conspiratorial channel.

\begin{table*}[]
\caption{
    Comparison of SERP-MS scores for top-10 search results with data from the reference study.
    The scores range from $\left \langle -1, 1 \right \rangle$, where -1 denotes a debunking and 1 a promoting stance towards the conspiracy.
    Only search results from queries that were executed both by the reference study and us are considered.
}
\small
\label{tab:hussein-comparison-search}
\begin{tabularx}{\textwidth}{lllp{16mm}X}
\toprule
% \multicolumn{5}{c}{Search results SERP-MS score comparison with the reference study}\\
Topic            & Hussein              & Ours   & Change & Inspection \\ \midrule

9/11             & -0.16                & -0.06 & No (n.s.d.)
% U=1200, p=0.7147
& Smaller changes that depend on search query.  \\

Chemtrails       & -0.2                 & -0.47 & No (n.s.d.)
% U=1322, p=0.6196
& Drop in promoting videos (from 45\% to 12\%) in 2 queries. \\

Flat earth       & -0.58                & -0.41 & No (n.s.d.)
% U=1500, p=0.0837
& 2 queries worsen a lot due to new content. Other queries improve. \\

Moon landing     & -0.6                 & -0.59 & No (n.s.d.)
% U=650	0.1446
& Smaller decrease in number of neutral and increase of debunking videos. \\

Anti-vaccination & -0.8                 & -0.63 & \textcolor{red}{Worse} (U=324,p=$1.3\mathrm{e}{-9}$)
& Drop in number of debunking and increase in number of neutral videos.  \\ \bottomrule
\end{tabularx}
\end{table*}

\begin{table*}[]
\caption{
    Comparison of normalized scores for up-next and top-5 recommendations with data from the reference study.
    Normalized scores range from $\left \langle -1, 1 \right \rangle$, where -1 denotes a debunking and 1 a promoting stance towards the conspiracy.
    Last 10 out of 20 watched videos in reference data are considered.
    Last 2 out of 40 watched videos in our data are considered.
}
\small
\label{tab:hussein-comparison-recommendations}
\begin{tabularx}{\textwidth}{llllX}
\toprule
% \multicolumn{5}{c}{Up-next + top-5 recommendations normalized score comparison with the reference study} \\
Topic            & Hussein & Ours   & Change  & Inspection \\ \midrule
9/11             & 0.14    & 0.26  & No (n.s.d.)                   & Similar distribution, more promoting videos. \\
Chemtrails       & 0.05    & 0.03  & No (n.s.d.)                   & More neutral results. \\
% U=2134.5, p=0.0624
Flat earth       & -0.16   & -0.15 & No (n.s.d.)                   & Similar distribution. \\
% U=1922.5, p=0.8321
Moon landing     & -0.08   & -0.32 & \textcolor{DarkGreen}{Better} (U=2954.5,p=$8\mathrm{e}{-6}$)   & More debunking videos. \\
Anti-vaccination & -0.28   & -0    & \textcolor{red}{Worse} (U=664,p=$1.6\mathrm{e}{-9}$)       & Less debunking videos, more neutral and promoting. \\ \bottomrule
\end{tabularx}
\end{table*}

%\paragraph{H1.1} --> moved to discussion
%This hypothesis stated that the tracked metrics improved since the reference study.
%We did not find a significantly different distribution of SERP-MS scores in overall search results.
%A single topic (anti-vaccination) showed a statistically significant difference.
%However, it did not agree with the hypothesis as the metric worsened due to more neutral videos.
%Recommendations showed significant differences across multiple topics but were not significantly different overall.
%A single topic improved normalized scores of recommendation in agreement with the hypothesis.
%Yet, two topics worsened their scores.
%We suspect the changes in search results and recommendations were influenced mostly by changes in content and do not show a significant improvement in fight against misinformation on the platform as stated in the hypothesis.

%Questions: how did Hussein evaluated recommendations, if not by SERP-MS (probably just normalized score?)

% \todo[inline]{offer some interpretations: do we suspect policy change? or influence of COVID? was it consistent across topics? methodological differences? ...}

\subsection{RQ2: What is the effect of watching debunking videos after the promoting phase?}

\begin{table*}[]
\caption{
Comparison of SERP-MS scores for top-10 search results in promoting and debunking phase of our experiment.
Three points are compared: start of promoting phase (S1), end of promoting phase (E1), end of debunking phase (E2).
}
\small
\label{tab:comparison-search}
\begin{tabularx}{\textwidth}{p{10mm}p{11mm}p{37mm}X}
\toprule
% \multicolumn{4}{c}{SERP-MS scores on top-10 search results} \\
Topic            & SERP-MS & Change & Inspection \\ \midrule
9/11             &
S1: -0.07 \newline E1: -0.06 \newline E2: -0.11
&
S1--E1: n.s.d. \newline
E1--E2: n.s.d. \newline
S1--E2: n.s.d.
% test towards end of phase 1: U=1108, p=0.1544
&
E2: More debunking videos in one query (30\% instead of 12\% at S1 and 11\% at E1 in query ``9/11'').
% test towards end of phase 2: U=1020, p=0.0494
\\ \hline
Chemtrails       &
S1: -0.45 \newline E1: -0.47 \newline E2: -0.49
&
S1--E1: n.s.d. \newline
E1--E2: n.s.d. \newline
S1--E2: \textcolor{DarkGreen}{better} (U=915,p=0.0097)
% test towards end of phase 1: U=1055.5, p=0.0886
&
E2: The ``Chemtrail'' search query showed an increase in number of debunking videos (from 66\% at S1 and 69\% at E1 to 80\%) and a decrease in promoting (from 10\% to 0\%).
% test from end of phase 1 to end of phase 2: U=1035, p=0.0659
% test from start end of phase 2: U=915, p=0.0097
\\ \hline
Flat earth       &
S1: -0.27 \newline E1: -0.41 \newline E2: -0.45
&
S1--E1: \textcolor{DarkGreen}{better} (U=762.5,p=0.0004) \newline
E1--E2: n.s.d. \newline
S1--E2: \textcolor{DarkGreen}{better} (U=704.5,p=0.0001)
&
E1: Change goes against expectations.
Promoting videos disappear in 3 search queries and decrease in another one (from 36\% to 30\%).
% test towards end of phase 1: U=762.5, p=0.0004
\newline
E2: Similar change as in E1 with a further decrease in promoting videos in one query (from 30\% to 22\%) and reordered videos in another.
% test towards end of phase 2: U=1112, p=0.1684
% test from start to end of phase 2: U=704.5, p=0.0001
\\ \hline
Moon landing     &
S1: -0.57 \newline E1: -0.57 \newline E2: -0.59
&
S1--E1: n.s.d. \newline
E1--E2: n.s.d. \newline
S1--E2: \textcolor{DarkGreen}{better} (U=900,p=0.0068)
% test towards end of phase 1: U=1195, p=0.3503
&
E2: Reordered search results in ``moan hoax'' query---debunking videos moved higher.
% test towards end of phase 2: U=960, p=0.0189
% test from start to end of phase 2: U=900, p=0.0068
\\ \hline
Anti-vacc. &
S1: -0.6 \newline E1: -0.63 \newline E2: -0.68
&
S1--E1: n.s.d. \newline
E1--E2: \textcolor{DarkGreen}{better} (U=699.5,p=0.0054) \newline
S1--E2: \textcolor{DarkGreen}{better} (U=641.5,p=0.0001)
% test towards end of phase 1: U=966.5, p=0.1179
&
E2: Increase in debunking videos across multiple queries (from 60\% at S1 and 61\% at E1 to 67\%).
% test towards end of phase 2: U=699.5, p=0.0054
% test from start to end of phase 2: U=641.5, p=0.0001
\\ \bottomrule
\end{tabularx}
\end{table*}

\begin{table*}[]
\caption{
Comparison of changes in average normalized scores for top-10 recommendations in promoting and debunking phase of our experiment.
Three points are compared: start of promoting phase (S1), end of promoting phase (E1), end of debunking phase (E2).
}
\small
\label{tab:comparison-recommendations}
\begin{tabularx}{\textwidth}{p{10mm}p{11mm}p{37mm}X}
\toprule
% \multicolumn{4}{c}{Normalized scores on top-10 recommendations} \\
Topic            & Score & Change & Inspection \\ \midrule
9/11             &
S1: 0.1 \newline E1: 0.42 \newline E2: 0.07
&
S1--E1: \textcolor{red}{worse} (U=45.5,p=$2.6\mathrm{e}{-5}$) \newline
E1--E2: \textcolor{DarkGreen}{better} (U=28,p=$2.9\mathrm{e}{-6}$) \newline
S1--E2: n.s.d.
&
E1: Number of promoting videos increased (from 14\% to 43\%) and neutral videos decreased (from 83\% to 56\%).
% test towards end of phase 1: U=45.5, p=0
\newline
E2: The numbers of promoting and neutral videos returned to levels comparable to start (13\% and 82\%).
% test towards end of phase 2: U=372, p=0
% test from start to end of phase 2: U=216.5, p=0.6514
\\ \hline
Chemtrails       &
S1: 0 \newline E1: 0.05 \newline E2: -0.15
&
S1--E1: n.s.d. \newline
E1--E2: \textcolor{DarkGreen}{better} (U=323, p=0.0006) \newline
S1--E2: \textcolor{DarkGreen}{better} (U=330, p=0.0002)
% test towards end of phase 1: U=195, p=0.886
&
E2: There is an increase in a number of debunking videos (from 0\% at S1 and 3\% at E1 to 19\%).
In return, we end up in a state that is better than at the start.
% test towards end of phase 2: U=323, p=0.0006
% test from start to end of phase 2: U=330, p=0.0002
\\ \hline
Flat earth       &
S1: -0.17 \newline E1: -0.06 \newline E2: -0.47
&
S1--E1: n.s.d. \newline
E1--E2: \textcolor{DarkGreen}{better} (U=375, p=$1.8\mathrm{e}{-6}$) \newline
S1--E2: \textcolor{DarkGreen}{better} (U=347, p=0.0001)
% test towards end of phase 1: U=141.5, p=0.1047
&
E2: Similar to the Chemtrails conspiracy, there is an increase in number of debunking videos (from 19\% at S1 and 16\% at E1 to 48\%).
% test towards end of phase 2: U=375, p=0
% test from start to end of phase 2: U=347, p=0.0001
\\ \hline
Moon landing     &
S1: -0.2 \newline E1: -0.4 \newline E2: -0.42
&
S1--E1: n.s.d. \newline
E1--E2: n.s.d. \newline
S1--E2: n.s.d.
% test towards end of phase 1: U=268, p=0.0642
&
E1: Mean normalized scores changes against expectation and improves (but not significantly).
% test towards end of phase 2: U=243, p=0.2395
% test from start to end of phase 2: U=290, p=0.0143
\\ \hline
Anti-vacc. &
S1: -0.1 \newline E1: 0.04 \newline E2: -0.37
&
S1--E1: \textcolor{red}{worse} (U=74.5,p=0.0008) \newline
E1--E2: \textcolor{DarkGreen}{better} (U=310,p=$2.5\mathrm{e}{-6}$) \newline
S1--E2: \textcolor{DarkGreen}{better} (U=307.5,p=0.0002)
% test towards end of phase 1: U=74.5, p=0.0016
&
E1: Increase in number of promoting videos (from 2\% to 13\%).
\newline
E2: Increase of debunking videos (from 12\% at S1 and 9\% at E1 to 37\%) and disappearance of promoting (from 2\% at S1 and 13\% at E1 to 0\%).
% test towards end of phase 2: U=310, p=0
% test from start to end of phase 2: U=307.5, p=0.0002
\\ \bottomrule 
\end{tabularx}
\end{table*}

Answering this question requires three comparisons:
\begin{enumerate}
    \item comparison of metrics between start of promoting phase (S1) and end of promoting phase (E1),
    \item comparison of metrics between end of promoting phase (E1) and end of debunking phase (E2),
    \item comparison of metrics between start of promoting phase (S1) and end of debunking phase (E2).
\end{enumerate}

\emph{Comparison (1)} shows changes in search results and recommendations after watching promoting videos (E1) compared to the start of the experiment (S1).
If there was a misinformation bubble created, we would expect the metrics to worsen due to watching promoting videos.
Regarding search results, the distribution of SERP-MS scores between S1 and E1 is indeed significantly different (MW U=34118.5, p-value=0.028).
However, the score actually improves---mean SERP-MS score changed from -0.39 (std 0.28) to -0.42 (std 0.3).
Table~\ref{tab:comparison-search} shows the change for individual topics.
Only the flat earth conspiracy shows significant differences and improved the SERP-MS score due to a decrease in promoting and an increase of debunking videos.
Top-10 recommendations also change their distribution of normalized scores significantly at E1 compared to S1 (MW U=4085, p-value=0.0397).
We observe that the mean normalized score worsens from -0.07 (std 0.24) to 0.01 (std 0.31).
Looking at individual topics in Table~\ref{tab:comparison-recommendations}, we can see that the change is significant in topics 9/11 and anti-vaccination that gain more promoting videos.
%In the moon landing topic, the score actually improves, but this change is not statistically significant.

\emph{Comparison (2)} relates the change in search results and recommendations between the end of promoting phase (E1) and the end of debunking phase (E2).
We expect the metrics would improve due to watching debunking videos, i.e., that we would observe misinformation bubble bursting.
However, SERP-MS scores in search results between E1 and E2 are not from statistically significantly different distributions, which is consistent with the fact that we did not observe misinformation bubble creation in search results in the first place.
Table~\ref{tab:comparison-search} shows that only a single topic---anti-vaccination---significantly changed its distribution and improved its mean score.
Nevertheless, we see minor improvements in SERP-MS scores also in other topics.
Top-10 recommendations show more considerable differences and their overall distribution is significantly different comparing E1 and E2 (MW U=7179.5, p-value=$1.8\mathrm{e}{-9}$).
Mean normalized score improves from 0.01 (std 0.31) to -0.27 (std 0.27).
Table~\ref{tab:comparison-recommendations} shows significantly different distributions for all topics except for moon landing conspiracy.
All topics show an improvement in normalized scores.
The 9/11 topic shows a decrease in promoting videos, while other topics show an increase in the number of debunking videos.

\emph{Comparison (3)} shows differences between the start (S1) and end of the experiment (E2).
We expect the metrics would improve due to watching debunking videos despite watching promoting videos before that.
The distribution of SERP-MS scores in search results is statistically significantly different when comparing S1 and E2 (MW U=36515, p-value=0.0002).
Overall, we see an improvement in mean SERP-MS score from -0.39 (std 0.28) to -0.46 (std 0.29).
In contrast with comparison (2), Table~\ref{tab:comparison-search} shows that all topics except 9/11 significantly changed their distributions.
All topics show an improvement according to our expectations.
The improvement is due to increases in debunking videos, decreases in promoting videos, or reordered search results in some search queries.
Similarly, top-10 recommendations at E2 come from a significantly different distribution than at S1 (MW U=6940.5, p-value=$2.9\mathrm{e}{-7}$).
Mean normalized score improves from -0.07 (std 0.24) to -0.27 (std 0.27).
Table~\ref{tab:comparison-recommendations} shows a significant difference in distributions for all topics except for 9/11 and moon landing conspiracies.
Mean normalized scores improve compared to S1 in all topics except for 9/11. %, where the score is slightly worse at E2 than at S1 (but the change is not statically significantly different).
Nevertheless, the numbers of promoting and neutral videos in 9/11 topic at E2 are comparable to S1.
Other topics show increases in the numbers of debunking videos.

\section{Discussion and Conclusions}

In the paper, we presented an audit of misinformation present in search results and recommendations on the video-sharing platform YouTube. To support reproducibility, we publish the collected data and source codes for the experiment.

We aimed at verifying a hypothesis that there is less misinformation present in both search results and recommendations after recent changes in YouTube policies~\cite{YouTube2020policies} (H1.1). The comparison was done against a study done in mid 2019 by Hussein et al.~\cite{Hussein2020}. We were interested, whether we could still observe the formation of misinformation bubbles after watching videos promoting conspiracy theories (H2.0). In contrast to the previous studies, we also examined bubble bursting behavior. Namely, we aimed to verify whether misinformation bubbles could be burst if we watched videos debunking conspiracy theories (H2.1). We also hypothesized that watching debunking videos (even after a previous sequence of promoting videos) would still decrease the amount of misinformation compared to the initial state with no watch history at the start of the study (H2.2).

Regarding hypothesis H1.1, we did not find a significantly different amount of misinformation in search results in comparison to the reference study. A single topic (anti-vaccination) showed a statistically significant difference. However, it did not agree with the hypothesis as the metric \emph{worsened} due to more neutral and less debunking videos. Recommendations showed significant differences across multiple topics but were not significantly different overall. A single topic (moon landing) improved normalized scores of recommendation in agreement with the hypothesis. Yet, the anti-vaccination topic worsened its scores.
We suspect the changes in search results and recommendations were influenced mostly by changes in content. Overall, our results did not show a significant improvement in the fight against misinformation on the platform, as stated in the hypothesis.

We did not observe the creation of misinformation filter bubbles in search results (H2.0) despite watching promoting videos. On the other hand, recommendations behaved according to our hypothesis, and their overall normalized scores worsened. Since there was no filter bubble creation effect in search results, we did not observe any bubble bursting effect there. Results did not show a statistically significant difference between the end of promoting phase and the end of the debunking phase. Only a single topic (anti-vaccination) showed a statistically significant difference and an improvement following the hypothesis H2.1. Recommendations showed more considerable differences that were statistically significant and confirmed the hypothesis. Lastly, we showed that watching debunking videos decreases the number of misinformation videos both in search results and recommendations, which confirms our hypothesis H2.2. We observed an improvement of SERP-MS scores in all topics except for one and an improvement of normalized scores for recommendations in most topics.

Based on our results, we can conclude that users, even with a watch history of promoting conspiracy theories, do not get enclosed in a misinformation filter bubble \emph{when they search} on YouTube. However, we do observe this effect in video recommendations with varying degrees depending on the topic. However, \emph{watching debunking videos helps in practically all cases} to decrease the amount of misinformation that the users see. Additionally, although we expected to see less misinformation than the previous studies reported, this was in general not the case. Worsening in the anti-vaccination topic was partially expected due to the COVID-19 pandemic. However, it is interesting that we also observed a worse situation with the 9/11 topic. In fact, this topic served as a sort of a gateway to misinformation videos on other topics.

A limitation of our results lies with the limited amount of topics that we investigated -- these did not include, for example, recent QAnon conspiracy and COVID-19 related conspiracies were present only through anti-vaccination narratives. However, our topics were explicitly selected to allow comparison with the reference study. Next, we included only a limited set of agent interactions with the platform (search and video watching). Real users also like or dislike videos, subscribe to channels, leave comments or click on the search results or recommendations. A more human-like bot simulation, with these interactions and possible inclusion of human biases bursting remains our future work.

Nevertheless, our audit showed that YouTube (similar to other platforms), despite their best efforts so far, can still promote misinformation seeking behavior to some extent. The results also motivate the need for independent continuous and automatic audits of YouTube and other social media platforms~\cite{Simko2021}, since we observed that the amount of misinformation in a topic could change over time due to endogenous as well as exogenous factors.

\begin{acks}
This research was partially supported by TAILOR, a project funded by EU Horizon 2020 research and innovation programme under GA No 952215.
\end{acks}


\begin{thebibliography}{33}
\bibitem{AbulFottouh2020}
Deena Abul-Fottouh, Melodie Yunju Song, and Anatoliy Gruzd. 2020. Examining algorithmic biases in YouTube’s recommendations of vaccine
videos. Int. Journal of Medical Informatics 140 (2020), 104175. https://doi.org/10.1016/j.ijmedinf.2020.104175

\bibitem{bakshy2015exposure}
Eytan Bakshy, Solomon Messing, and Lada A. Adamic. 2015. Exposure to ideologically diverse news and opinion on Facebook. Science 348, 6239
(2015), 1130–1132.

\bibitem{bessi2016personality}
Alessandro Bessi. 2016. Personality traits and echo chambers on facebook. Computers in Human Behavior 65 (2016), 319–324.

\bibitem{Covington2016}
Paul Covington, Jay Adams, and Emre Sargin. 2016. Deep Neural Networks for YouTube Recommendations. In Proc. of the 10th ACM Conference on
Recommender Systems (RecSys ’16). ACM, New York, NY, USA, 191–198. https://doi.org/10.1145/2959100.2959190

\bibitem{del2016spreading}
Michela Del Vicario, Alessandro Bessi, Fabiana Zollo, Fabio Petroni, Antonio Scala, Guido Caldarelli, H Eugene Stanley, and Walter Quattrociocchi.
2016. The spreading of misinformation online. Proc. of the National Academy of Sciences 113, 3 (2016), 554–559.


\bibitem{festinger1957theory}
Leon Festinger. 1957. A theory of cognitive dissonance. Vol. 2. Stanford university press.


\bibitem{Hannak2013}
Aniko Hannak, Piotr Sapiezynski, Arash Molavi Kakhki, Balachander Krishnamurthy, David Lazer, Alan Mislove, and Christo Wilson. 2013.
Measuring Personalization of Web Search. In Proc. of the 22nd International Conference on World Wide Web (WWW ’13). ACM, New York, NY, USA,
527–538. https://doi.org/10.1145/2488388.2488435


\bibitem{Hussein2020}
Eslam Hussein, Prerna Juneja, and Tanushree Mitra. 2020. Measuring Misinformation in Video Search Platforms: An Audit Study on YouTube. Proc.
ACM Hum.-Comput. Interact. 4, CSCW1, Article 048 (May 2020), 27 pages. https://doi.org/10.1145/3392854

\bibitem{Juneja2021}
Prerna Juneja and Tanushree Mitra. 2021. Auditing E-Commerce Platforms for Algorithmically Curated Vaccine Misinformation. In Proc. of the 2021
CHI Conference on Human Factors in Computing Systems (CHI ’21). https://doi.org/10.1145/3411764.3445250 arXiv:2101.08419


\bibitem{Le2019}
Huyen Le, Andrew High, Raven Maragh, Timothy Havens, Brian Ekdale, and Zubair Shafiq. 2019. Measuring political personalization of Google
news search. In Proc. of the World Wide Web Conference (WWW ’19). 2957–2963. https://doi.org/10.1145/3308558.3312504

\bibitem{RePEc:now:jlqjps:100.00016037}
Ben Lockwood. 2017. Confirmation Bias and Electoral Accountability. Quarterly Journal of Political Science 11, 4 (February 2017), 471–501.
https://doi.org/10.1561/100.00016037

\bibitem{Metaxa2019}
Danaë Metaxa, Joon Sung Park, James A. Landay, and Jeff Hancock. 2019. Search Media and Elections: A Longitudinal Investigation of Political
Search Results. Proc. ACM Hum.-Comput. Interact. 3, CSCW, Article 129 (Nov. 2019), 17 pages. https://doi.org/10.1145/3359231


\bibitem{mutz2011communication}
Diana C. Mutz and Lori Young. 2011. Communication and public opinion: Plus ça change? Public opinion quarterly 75, 5 (2011), 1018–1044.


\bibitem{Papadamou2020}
Kostantinos Papadamou, Savvas Zannettou, Jeremy Blackburn, Emiliano De Cristofaro, Gianluca Stringhini, and Michael Sirivianos. 2020. "It is just
a flu": Assessing the Effect of Watch History on YouTube’s Pseudoscientific Video Recommendations. arXiv:2010.11638 [cs.CY]


\bibitem{pariser2011filter}
Eli Pariser. 2011. The filter bubble: What the Internet is hiding from you. Penguin UK.


\bibitem{pecher2021fireant}
Branislav Pecher, Ivan Srba, Robert Moro, Matus Tomlein, and Maria Bielikova. 2021. FireAnt: Claim-Based Medical Misinformation Detection
and Monitoring. In Proc. of the Joint European Conference on Machine Learning and Knowledge Discovery in Databases (ECML PKDD ’20). 555–559.
https://doi.org/10.1007/978-3-030-67670-4\_38


\bibitem{Ribeiro2020}
Manoel Horta Ribeiro, Raphael Ottoni, Robert West, Virgílio A. F. Almeida, and Wagner Meira. 2020. Auditing Radicalization Pathways on YouTube.
ACM, New York, NY, USA, 131–141. https://doi.org/10.1145/3351095.3372879


\bibitem{Robertson2018}
Ronald E. Robertson, David Lazer, and Christo Wilson. 2018. Auditing the personalization and composition of politically-related search engine
results pages. Proc. of the World Wide Web Conference (WWW ’18), 955–965. https://doi.org/10.1145/3178876.3186143

\bibitem{Sandvig2014Audits}
Christian Sandvig, Kevin Hamilton, Karrie Karahalios, and Cedric Langbort. 2014. Auditing algorithms: Research methods for detecting discrimination
on internet platforms. Data and discrimination: converting critical concerns into productive inquiry 22 (2014), 4349–4357.


\bibitem{Silva2020}
Márcio Silva, Lucas Santos de Oliveira, Athanasios Andreou, Pedro Olmo Vaz de Melo, Oana Goga, and Fabricio Benevenuto. 2020. Facebook Ads
Monitor: An Independent Auditing System for Political Ads on Facebook. In Proc. of The Web Conference (WWW ’20). ACM, New York, NY, USA,
224–234. https://doi.org/10.1145/3366423.3380109


\bibitem{simko2021study}
Jakub Simko, Patrik Racsko, Matus Tomlein, Martina Hanakova, Robert Moro, and Maria Bielikova. 2021. A study of fake news reading and
annotating in social media context. New Review of Hypermedia and Multimedia (2021), 1–31.

\bibitem{Simko2021}
Jakub Simko, Matus Tomlein, Branislav Pecher, Robert Moro, Ivan Srba, Elena Stefancova, Andrea Hrckova, Michal Kompan, Juraj Podrouzek,
and Maria Bielikova. 2021. Towards Continuous Automatic Audits of Social Media Adaptive Behavior and Its Role in Misinformation Spreading.
In Adjunct Proc. of the 29th ACM Conference on User Modeling, Adaptation and Personalization (UMAP ’21). ACM, New York, NY, USA, 411–414.
https://doi.org/10.1145/3450614.3463353

\bibitem{Spinelli2020}
Larissa Spinelli and Mark Crovella. 2020. How YouTube Leads Privacy-Seeking Users Away from Reliable Information. In Adjunct Publication of the
28th ACM Conference on User Modeling, Adaptation and Personalization. ACM, New York, NY, USA, 244–251. https://doi.org/10.1145/3386392.3399566


\bibitem{srba2019monant}
Ivan Srba, Robert Moro, Daniela Chuda, Maria Bielikova, Jakub Simko, Jakub Sevcech, Daniela Chuda, Pavol Navrat, and Maria Bielikova. 2019.
Monant: Universal and Extensible Platform for Monitoring, Detection and Mitigation of Antisocial Behavior. In Proc. of Workshop on Reducing
Online Misinformation Exposure (ROME ’19). 1–7.


\bibitem{sunstein1999law}
Cass R Sunstein. 1999. The law of group polarization. University of Chicago Law School, John M. Olin Law and Economics Working Paper 91 (1999).


\bibitem{Tranberg2020Dataethics}
Pernille Tranberg, Gry Hasselbalch, Catrine S. Byrne, and Birgitte K. Olsen. 2020. DATAETHICS – Principles and Guidelines for Companies, Authorities
and  Organisations. Dataethics.eu. https://spintype.com/book/dataethics-principles-and-guidelines-for-companies-authorities-organisations


\bibitem{vaidhyanathan2018antisocial}
Siva Vaidhyanathan. 2018. Antisocial media: How Facebook disconnects us and undermines democracy. Oxford University Press.


\bibitem{YouTube2020policies}
YouTube. 2020. Managing harmful conspiracy theories on YouTube. https://blog.youtube/news-and-events/harmful-conspiracy-theories-youtube/


\bibitem{Zannettou2019}
Savvas Zannettou, Michael Sirivianos, Jeremy Blackburn, and Nicolas Kourtellis. 2019. The Web of False Information: Rumors, Fake News, Hoaxes,
Clickbait, and Various Other Shenanigans. Journal of Data and Information Quality (2019), 1–37. https://doi.org/10.1145/3309699 arXiv:1804.03461


\bibitem{Zhao2019}
Zhe Zhao, Lichan Hong, Li Wei, Jilin Chen, Aniruddh Nath, Shawn Andrews, Aditee Kumthekar, Maheswaran Sathiamoorthy, Xinyang Yi, and Ed
Chi. 2019. Recommending what video to watch next: A multitask ranking system. In Proc. of the 13th ACM Conference on Recommender Systems
(RecSys ’19). ACM, 43–51. https://doi.org/10.1145/3298689.3346997


\bibitem{Zhou2020}
Xinyi Zhou and Reza Zafarani. 2020. A Survey of Fake News: Fundamental Theories, Detection Methods, and Opportunities. Comput. Surveys 53, 5
(Dec. 2020). https://doi.org/10.1145/3395046 arXiv:1812.00315


\bibitem{journals/corr/ZolloNVBMSCQ15}
Fabiana Zollo, Petra Kralj Novak, Michela Del Vicario, Alessandro Bessi, Igor Mozetic, Antonio Scala, Guido Caldarelli, and Walter Quattrociocchi.
2015. Emotional Dynamics in the Age of Misinformation. CoRR (2015). http://dblp.uni-trier.de/db/journals/corr/corr1505.html\#ZolloNVBMSCQ15


\bibitem{zuboff2019age}
Shoshana Zuboff. 2019. The Age of Surveillance Capitalism: The Fight for a Human Future at the New Frontier of Power. Profile Books.



\end{thebibliography}
\end{document}